\documentclass[aps,prb,twocolumn,groupedaddress,showpacs]{revtex4}
\usepackage{epsfig}
\usepackage[dvipsnames,usenames]{color}

\emergencystretch=\maxdimen
\begin{document}

\title{Phase Stability in the Two dimensional Anisotropic
Boson Hubbard Hamiltonian}

\author{T. Ying$^{1,2}$, G.G.~Batrouni$^{3,4,5}$,
V.G. Rousseau$^6$, M. Jarrell$^6$,
J. Moreno$^6$, X.D.~Sun$^1$, and R.T.~Scalettar$^2$}

\affiliation{$^1$Department of Physics, Harbin Institute of
Technology, Harbin 150001, China}

\affiliation{$^2$Physics Department, University of California, Davis,
California 95616, USA}

\affiliation{$^3$INLN, Universit\'e de Nice--Sophia Antipolis, CNRS;
1361 route des Lucioles, 06560 Valbonne, France}

\affiliation{$^4$Institut Universitaire de France}

\affiliation{$^5$Centre for Quantum Technologies, National
University of Singapore; 2 Science Drive 3 Singapore 117542}

\affiliation{$^6$Department of Physics and Astronomy,
Louisiana State University, Baton Rouge, Louisiana 70803, USA}

\begin{abstract}
The two dimensional square lattice hard-core boson Hubbard model with
near neighbor interactions has a `checkerboard' charge density wave
insulating phase at half-filling and sufficiently large intersite
repulsion.  When doped, rather than forming a supersolid phase in which
long range charge density wave correlations coexist with a condensation
of superfluid defects, the system instead phase separates.  However, it
is known that there are other lattice geometries and interaction
patterns for which such coexistence takes place.  In this paper we
explore the possibility that anisotropic hopping or anisotropic near
neighbor repulsion might similarly stabilize the square lattice
supersolid.  By considering the charge density wave structure factor and
superfluid density for different ratios of interaction strength and
hybridization in the $\hat x$ and $\hat y$ directions, we conclude that phase
separation still occurs.
\end{abstract}


\maketitle

\section{INTRODUCTION}

In fermionic systems the competition between phase separation and
(non-phonon mediated) superconductivity has been a long-standing
and active area of research.
On the one hand, Cooper pair formation requires an effective attractive
interaction, but on the other hand this attraction might `go overboard' and
lead to the agglomeration of larger clusters of particles.
This latter possibility is especially likely in commonly studied models
of superconductivity like the Hubbard and $t$-$J$ Hamiltonians
\cite{emery90,marder90,hellberg97,riera89,dagotto90,fye90,chang08}
which do not include a long-range Coulomb interaction that
opposes phase separation.

A very similar set of issues arises in the interplay of Bose-Einstein
condensation (and the closely related phenomenon of superfluidity)
with charge order in bosonic systems, where the possibility of
coexistence of diagonal and off-diagonal long range order is termed a
`supersolid' (SS) phase \cite{supersolid}.  Bosonic systems are
commonly studied in the continuum\cite{pollock87,khairallah95},
motivated by liquid $^4$He.  However, as with fermions, issues such as
supersolidity can also be addressed within a tight-binding lattice
model governed by the boson Hubbard Hamiltonian\cite{fisher89},
\begin{eqnarray}
\label{Hamiltonian}
{\hat\mathcal H}&=&-\sum_{ i\,j }
t_{ij}
(a_{i}^\dagger a_{j}^{\phantom{\dagger}} +
a_{j}^\dagger a_{i}^{\phantom{\dagger}} )
+ \frac12 \sum_{i} U_{i} n_{i} (n_{i} -1 )
\nonumber
\\
&+&\sum_{ i\,j }
V_{ij}
n_{ i} n_{j}
- \sum_{i} \mu_{i} n_{i}
\label{ham}
\end{eqnarray}
Here $a_{i}^\dagger$, $a_{i}^{\phantom{\dagger}}$, and
$n_{i}^{\phantom{\dagger}}$ are boson creation, destruction, and
number operators.  $t_{ij}$ is an intersite hopping between sites
${i}$ and ${j}$.  In this paper we consider a square lattice of linear
size $L$ and total number of sites $N=L^2$.  $U_{i}$ is an on-site
repulsion, and $V_{ij}$ is an intersite repulsion.  The chemical
potential $\mu$ controls the filling of the lattice.  Most studies of
the boson-Hubbard Hamiltonian have considered the case of
translationally invariant and short range interactions $U_{i} = U$,
$V_{ij}=V_1(V_2)$ for near neighbor (next near neighbor) pairs ${ij}$,
and near neighbor hopping $t_{ij}=t$.  More complex choices for the
Hamiltonian parameters are increasingly being studied, most notably
situations in which the chemical potential models a quadratic trap to
confine the bosons
\cite{kashurnikov02,batrouni02,wessel04,batrouni08}.

In the uniform interaction, hard-core limit, $U=\infty$, a site can be
either empty or singly occupied.  When a near neighbor, $V_1$, is
turned on at half-filling, the bosons begin to occupy only one of the
two sublattices into which the square lattice is divided, owing to its
bipartite character.  For $V_1$ sufficiently large and at low
temperature, this tendency becomes sufficiently pronounced to result
in long range order: the boson occupations alternate in a checkerboard
pattern, with a charge ordering vector $q=(\pi,\pi)$ and charge
correlations extend to large separation.  Doping away from
half-filling creates defects in this charge density wave (CDW) phase.
If the defect concentration is sufficiently low, it is possible that
long range charge order is not destroyed, even though the defects can
move through the ordered lattice and form a superfluid.  In this way a
supersolid would be realized.

However, it was shown\cite{batrouni95,batrouni00} that this
coexistence does not occur for hardcore bosons on the 2D square
lattice with near neighbor interactions.  Rather, there is a
thermodynamic instability to phase separation in which different
spatial regions are superfluid and charge ordered.  Despite the
absence of simultaneous diagonal and off-diagonal long range order in
this most simple scenario, other lattice geometries have been shown to
exhibit supersolid phases, and, indeed, by now there is a considerable
numerical literature as a function of lattice geometry,
dimensionality, filling and interaction
\cite{otterlo05,goral02,wessel05,boninsegni05,sengupta05,batrouni06,scarola06,yi07,suzuki07,gan07,dang08,yamamoto09,danshita09,pollet10,capogrossosansone10,xi11,yamamoto12,jiang12,ohgoe12a,ohgoe12b}. More
exotic processes within the Hamiltonian, such as ring exchange, are
also known to play a potential role in supersolidity
\cite{ceperley04,rousseau05}.

One of the lines of investigation into supersolidity has pursued the
effect of anisotropy.  Thus, although the checkerboard supersolid with
$q=(\pi,\pi)$ ordering wavevector is unstable, it was also shown that
a striped supersolid is stable\cite{batrouni00}.  Such a phase is
caused by a next-near neighbor interaction $V_2$ which drives charge
order at wavevector $q=(\pi,0)$ or $(0,\pi)$ which is then doped to
allow for mobile defects.  In this case the underlying Hamiltonian is
isotropic, but a spontaneous breaking of the $Z_2$ symmetry associated
with order along the $\hat x$ or $\hat y$ directions occurs.  Recent
studies have also shown that symmetry breaking in the Hamiltonian
itself can give rise to striped supersolids.  In particular, Chan {\it
  et al.}\cite{chan10} argued that an attractive interaction in one
spatial direction and a repulsive interaction in the other will cause
supersolid formation.

These known instances of stability of the striped supersolid, as well
as the general delicacy of the energy balance involved in whether
phase separation occurs or not\cite{fye90}, suggest that a
checkerboard supersolid might be stabilized by anisotropic hoppings
$t_x \neq t_y$ or interactions $V_x \neq V_y$ in the $\hat x$ and
$\hat y$ directions.  This is the topic of the present paper.  Stable
supersolid phases have already been found for soft-core bosons in one,
two and three dimensions
\cite{sengupta05,batrouni06,ohgoe12a,ohgoe12b,batrouni13}. When the
contact repulsion, $U$, is large compared to the near neighbor
interaction, $V_1$, (the mean field value is $U>4V_1$ in two
dimensions), the system behaves as in the hard core case and phase
separates when doped away from half filling. When $V_1$ dominates,
$U<4V_1$, and the system is doped above half filling, the extra bosons
will go on already occupied sites. These doubly occupied defects then
delocalize and yield a supersolid phase. On the other hand, if the
system is doped below half-filling, it will typically undergo phase
separation. Remarkably, however, there is a very narrow parameter
range where the system will become supersolid when doped below half
filling \cite{ohgoe12a,ohgoe12b}.  This history of successful searches
for supersolidity in soft-core models prompts us to focus here on the
hard-core case.

After introducing our Quantum Monte Carlo methodology and the
observables used to characterize the phases, we show results
indicating that supersolids do not emerge from a checkerboard charge
ordered pattern driven by near-neighbor interactions, even when those
interactions or the hoppings are made anisotropic.  The dependences of
the superfluid density $\rho_s$ and the CDW structure factor $S(q)$ on
$V_{1x}/V_{1y}$ and $t_x/t_y$ are calculated.

\section{Methodology}

Our computational approach is the stochastic Green function (SGF)
approach\cite{SGF,directedSGF,explainSGF}.  SGF is a finite
temperature Quantum Monte Carlo algorithm which works in continuous
imaginary time (avoiding the need to extrapolate to zero
discretization mesh of the inverse temperature $\beta$).  Here we use
a recent new formulation of the method which includes global
space-time updates\cite{moreSGF} to explore phase space efficiently
and to sample the winding and hence superfluid density.  We use a
canonical formulation which allows us to work with fixed particle
number, although the method can also be implemented in the grand
canonical ensemble.  Note that even within a canonical implementation,
it is still possible to obtain the chemical potential as a finite
difference of ground state energies at adjacent particle numbers,
$\mu(n) = E_0(n+1) - E_0(n)$.

We will characterize the phases of the Hamiltonian Eq.~\ref{ham}
by examining the energy, $E=\langle H \rangle$ and also the
superfluid density $\rho_s$ and CDW structure factor
$S(q)$.  The superfluid density is obtained using the procedure
described by Pollock and Ceperley\cite{pollock87}, in which
$\rho_s$ is expressed in terms of the winding $W$ of bosonic world lines
across a lattice edge.
\begin{eqnarray}
\rho_{sx} = \frac{\langle W_x^2 \rangle}{2 \beta \, t_x}
\hskip0.70in
\rho_{sy} = \frac{\langle W_y^2 \rangle}{2 \beta \, t_y}
\label{rhos}
\end{eqnarray}
Here we have allowed for the possibilty of inequivalent spatial
directions, and measure the $x$ and $y$ superfluid densities separately.

The CDW structure factor is the Fourier transform of the
density-density correlation function.
\begin{eqnarray}
S(q) = \frac{1}{N^2} \sum_{jl} e^{i q\cdot l}
\langle n_{j} n_{j+l} \rangle
\label{sofq}
\end{eqnarray}
Here $N$ is the number of sites.
In order to establish long range order, both $\rho_s$ and
$S(q)$ must be extrapolated to the $N\rightarrow \infty$ limit
via finite size scaling.

The choice of how to compare anisotropic lattices with isotropic ones
is somewhat arbitrary.  Here we have chosen to keep $t_x+t_y$ fixed
as we vary $t_x$ away from $t_y$, rather than, for example, fixing $t_y$
and reducing $t_x$ to zero.  Our rationale is that this preserves the overall
noninteracting bandwidth $W=4(t_x+t_y)$.
Similarly, when we vary
the interactions we do so by fixing the sum $V_{1x} + V_{1y}$.
The motivation behind this choice is that in the strong coupling limit
the energy cost to
add a particle to a perfect checkerboard solid, $2(V_{1x}+V_{1y})$
is preserved.  That is, the ``base of the insulating
lobes"\cite{fisher89} at $t=0$ is the same for all degrees of
anisotropy.

\begin{figure}[htp]
\epsfig{figure=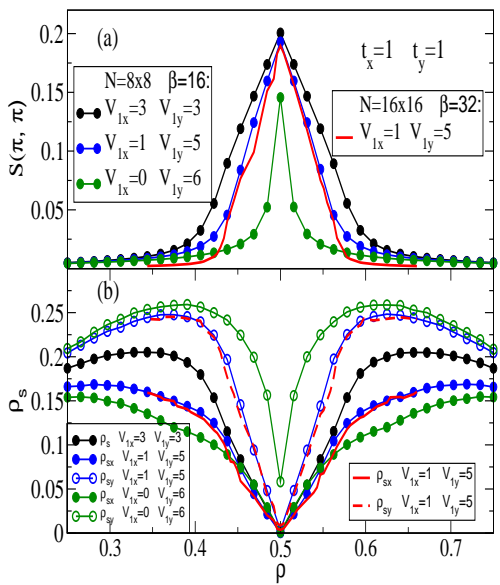,width=8.0cm,height=8.0cm,angle=-90,clip}
\caption{(Color online) Checkerboard structure factor (a) and
  superfluid density (b) as functions of filling, comparing
  anisotropic near neighbor interaction strengths $V_{1x}, V_{1y}$
  with isotropic ones.  $S(\pi,\pi)$ is relatively unaffected by
  shifting $V_{1x}$ away from $V_{1y}$: the data for $V_{1x}=1,
  V_{1y}=5$ declines somewhat more rapidly with doping away from
  $\rho=1$ but looks qualitatively quite similar.  Only when
  $V_{1x}=0$ does the structure factor change behavior significantly.
  Similar remarks apply to the superfluid density.  Superflow
  $\rho_{sy}$ in the $\hat y$ direction is larger than that in the
  $\hat x$ direction.  Most data are for 8x8 lattices at $\beta=16$,
  but lines without symbols are for 16x16 lattices at $\beta=32$ and
  show that finite spatial lattice and temperature effects are
  minimal.  Here and in subsequent figures error bars are smaller than
  the symbol size and hence are not shown.}
\label{fig1}
\end{figure}

\section{Results}

\subsection{Interaction Anisotropy}

We begin by considering $t_x=t_y$ but $V_{1x} \neq V_{1y}$.
Figure~\ref{fig1} shows the checkerboard structure factor $S(\pi,\pi)$
and superfluid density $\rho_s$ as functions of filling $\rho$.  In
Figure~\ref{fig1}(a), $S(\pi,\pi)$ is robust to anisotropy, declining
only modestly from its value at $V_{1x}=3, V_{1y}=3$ to its value at
$V_{1x}=1, V_{1y}=5$.  However, when $V_{1x}=0$, $S(\pi,\pi)$ is
greatly altered.  In this limit, if the hopping $t$ is neglected, the
lattice decomposes into independent chains each with density order at
$q_y=\pi$ but arbitrary $q_x$.  Figure~\ref{fig1} indicates that
nonzero $V_{1x}$ locks these chains into place spatially (selecting
out $q_x=\pi$ as well).  This is likely due to the enhanced kinetic
energy when hard core bosons in one chain find empty sites on
neighboring chains.  This robustness of checkerboard order in the
$V_{1x}=0$ limit is confined to $\rho=0.5$.  The structure factor
falls off much more abruptly with doping than when $V_{1x}$ remains
nonzero.

\begin{figure}[htp]
\epsfig{figure=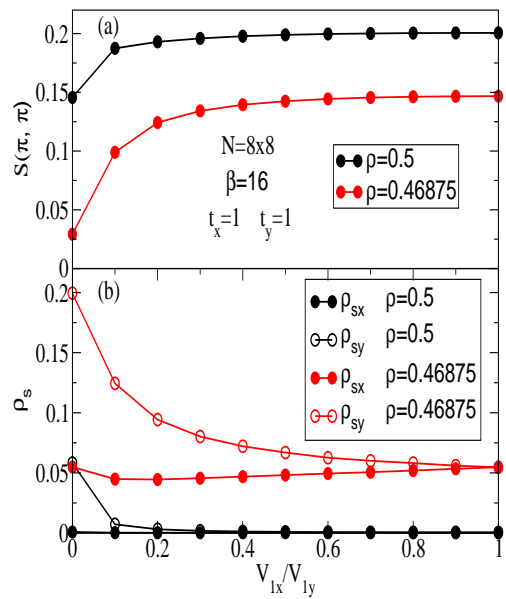,width=8.0cm,height=8.0cm,angle=-90,clip}
\caption{(Color online) (a) $S(\pi,\pi)$ vs $V_{1x}/V_{1y}$ at half-filling
  $\rho=0.5$ and lightly doped $\rho=15/32$.  The sum $V_{1x}+V_{1y}=6$
is fixed.  The flatness of the data
  except in the vicinity of $V_{1x}/V_{1y}=0$ emphasizes the minor
  effects of anisotropic interations, especially at half-filling.  (b)
  $\rho_{sx}$ and $\rho_{sy}$ vs $V_{1x}/V_{1y}$ at the same two
  densities.  The system is insulating for $\rho=0.5$ except strictly
  in the $V_{1x}=0$ limit.  In the doped case, $\rho_{sx}$ shows
  little change with anisotropy, while $\rho_{sy}$ increases sharply
  near $V_{1x}=0$.  Spatial lattice size is 8x8 and inverse
  temperature $\beta=16$.
}
\label{fig2}
\end{figure}

The behavior of $\rho_s$ in Figure~\ref{fig1}(b) is also largely
unaffected by anisotropy.  The largest change is an enhancement of
$\rho_s$ in the vicinity of the insulating solid phase at
half-filling.  Surprisingly, the details of the choice of interaction
strength become unimportant at low density (and high density due to
particle-hole symmetry).  The $x$ and $y$ components of the superfluid
response, $\rho_{sx}$ and $\rho_{sy}$, have the same qualitative
dependence on density $\rho$, although for the largest anisotropy
$V_{1x}=0$, $\rho_{sy}$ is roughly twice $\rho_{sx}$ for all
densities.  A precise picture of the interplay of the anisotropies in
charge correlations and superfluidity is not clear.  Since $V_{1x} <<
V_{1y}$ it is more likely that $\hat x$ neighbors will both be
occupied (or empty) than $\hat y$ neighbors.  Strings of adjacent
occupied sites of hard-core bosons would seemingly block motion
perpendicular to them (and hence suppress $\rho_{sy}$).  Meanwhile,
strings of adjacent empty sites would offer channels for an enhanced
$\rho_{sy}$.  A complicating feature of any attempt to build such a
`blocking/channel' picture is the fact that while the hard-core nature
of the bosons completely precludes certain types of motion, the strong
$V_{1y}$ interactions will also profoundly affect the pattern of
bosonic flow.  The small non-zero value of $\rho_{sy}$ at $V_{1x}=0$
and half-filling is a finite lattice effect.  (See Fig.~\ref{fig4}.)

\begin{figure}[htp]
\epsfig{figure=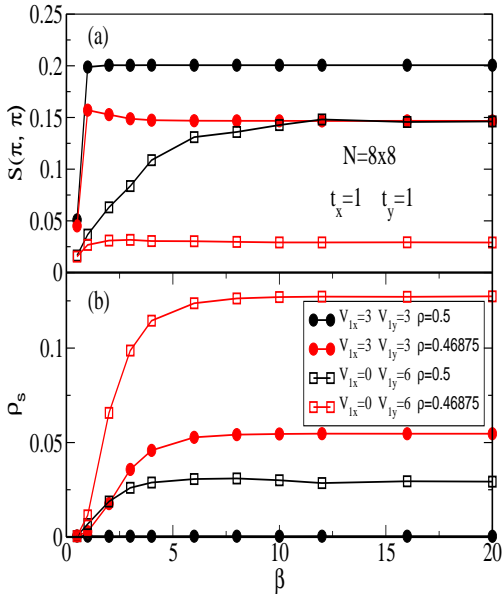,width=8.0cm,height=8.0cm,angle=-90,clip}
\caption{ \label{Tdenpendence} (Color online)The temperature dependence of (a)
  structure factor $S(\pi,\pi)$ and (b) superfluid density $\rho_s$
  (averaged over the $\hat x$ and $\hat y$ directions)
  are shown. We see that for $\beta \geq 10$, these quantities take
  their ground state values for the parameters we are using.  }
\end{figure}

The detailed evolution of the CDW structure factor and superfluid
densities with interaction anisotropy $V_{1x}/V_{1y}$ is given in
Fig.~\ref{fig2}.  Panel (a) emphasizes that $S(\pi,\pi)$ is generally
robust to anisotropy at half-filling $\rho=0.5$, but can change
dramatically for sufficiently small $V_{1x}/V_{1y}$ when the system is
(slightly) doped to $\rho=15/32$.  Panel (b) indicates similarly that
$\rho_s$ is much more affected by anisotropy in the presence of
vacancies.  Bosons flow more freely in the direction perpendicular to
that for which the repulsive interaction is smaller.  This probably
represents the low energy cost of lines of bosons which have strong
charge ordering in the $\hat y$ direction slipping past each other due
to the minimal energy cost in $V_{1x}$.  Of course such a picture is
a bit naive in the present case when the system undergoes phase
separation and consists of a mixture of superfluid and CDW regions.

\begin{figure}[htp]
\epsfig{figure=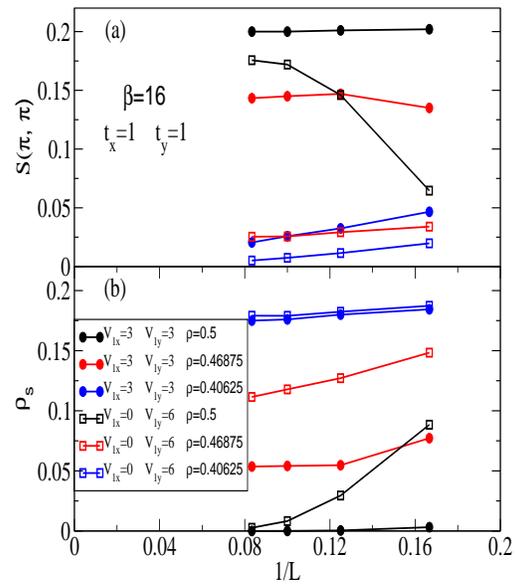,width=8.0cm,height=8.0cm,angle=-90,clip}
\caption{ \label{fig4} (Color online)The lattice size dependences of (a) structure
  factor $S(\pi,\pi)$ and (b) superfluid density $\rho_s$ are shown.
  Solid circles are the isotropic case $V_{1x}=V_{1y}=3$.  The system retains
  checkerboard CDW order when the doping is small ($\rho=15/32$) but
  not for $\rho=13/32$.  Open squares are the extreme anisotropic case
  $V_{1x}=0$.  In that case there is order only at half-filling.
  $\rho_s$ is larger when the interactions are anisotropic, but the
  finite size scaling indicates that it still vanishes at half-filling
in the thermodynamic limit even when $V_{1x}=0$.  }
\end{figure}

Figure~\ref{fig1}(a,b) contained some comparison of data on different
lattice sizes.  Fig.~\ref{Tdenpendence} and Fig.~\ref{fig4} show the
effects of temperature $T=1/\beta$ and $L$ more systematically.
Fig.~\ref{Tdenpendence} establishes that the system is in the ground
state when $\beta \approx 10$, validating the choice of $\beta=16$
used to access the ground state in Fig.~\ref{fig1}.  The approach to
the low temperature limit is somewhat more gradual in the extreme
anisotropic case $V_{1x}=0$.  Fig.~\ref{fig4} allows an extrapolation
to the thermodynamic limit $1/L \rightarrow 0$.  As at
half-filling, for a doping $\rho =15/32$ the CDW structure factor
$S(\pi,\pi)$ is nonzero at large $L$.  Further doping to $\rho =13/32$
destroys long range density order.  The anisotropic system is more
sensitive to doping than when $V_{1x}=V_{1y}$.

\begin{figure}[htp]
\epsfig{figure=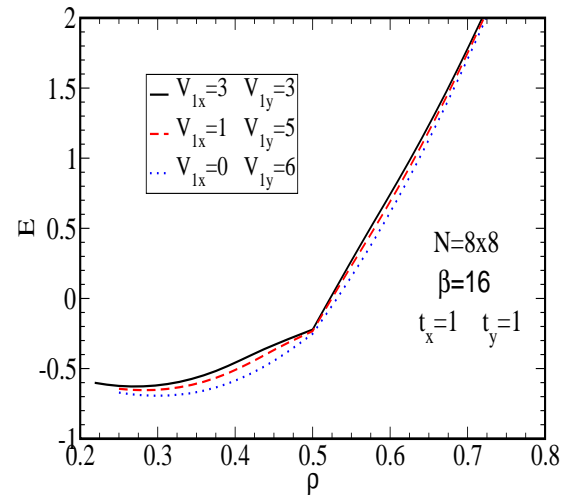,width=8.0cm,height=8.0cm,angle=-90,clip}
\caption{ \label{compare-EanisoV}
(Color online) Energy $E$ as a function of density $\rho$ exhibits the characteristic
signal of phase separation $d^2E/d\rho^2 <0$ as long as $V_{1x}$ remains
non-zero.  The kinks at half-filling imply an abrupt jump in chemical
potential, that is, the presence of a non-zero charge gap.
}
\end{figure}

Figures~\ref{fig1},~\ref{Tdenpendence}, \ref{fig4}
establish the possibility of a supersolid phase: Non-zero $\rho_s$
coexists with large $S(\pi,\pi)$ in the vicinity of half-filling.
However, it is necessary to distinguish a true supersolid phase, in
which these orders coexist as a single stable thermodynamic phase,
from a lattice in which both orders are present at separate spatial
locations.  This phase separation is most easily detected by looking
for an anomalous negative curvature, $d^2E/d\rho^2 < 0$, in the
thermodynamics.  Figure~\ref{compare-EanisoV} shows that such negative
curvature is present just below $\rho=0.5$, unless $V_{1x}=0$.

\begin{figure}[htp]
\epsfig{figure=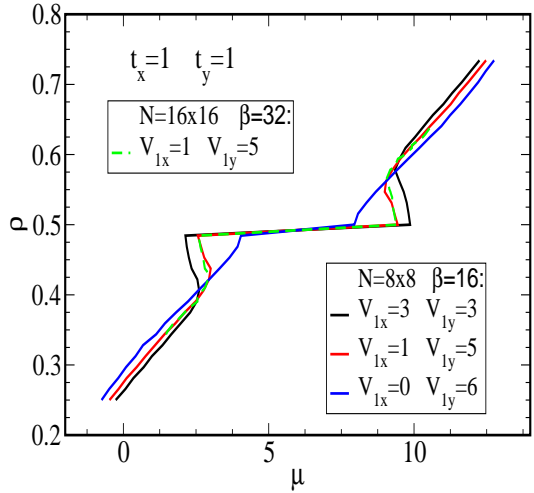,width=8.0cm,height=8.0cm,angle=-90,clip}
\caption{ \label{fig6} (Color online) Phase separation is indicated by
  a chemical potential $\mu$ which decreases as $\rho$ approaches
  $\rho=0.5$ from below.  This phenomenon, previously established in
  the isotropic case, is still present with anisotropy, except in the
  extreme limiting case $V_{1x}=0$. Results for 8x8 lattices at
  $\beta=16$ and 16x16 lattices at $\beta=32$ show finite-size and
  finite temperature effects are negligible both for the size of the
  Mott gap and the negative curvature (phase separation).  }
\end{figure}

The jump in slope $\mu = dE/d\rho$ at $\rho=0.5$ is the charge gap
$\Delta_c$.  Figure~\ref{fig6} shows the density
$\rho$ as a function of chemical potential $\mu$.  The kink at
half-filling in Fig.~\ref{compare-EanisoV} is reflected in a plateau
in the density.  As commented earlier, fixing $V_{1x}+V_{1y}$ makes
the length of this plateau $\Delta_c=2(V_{1x}+V_{1y})$ the same for
all degrees of anisotropy in the strong coupling ($t\rightarrow 0$)
limit.  Figure~\ref{fig6} indicates that finite
hopping $t=1$ reduces $\Delta_c$ from the $t=0$ value $\Delta_c=12$ to
$\Delta_c \sim 8$.  There is only a relatively small reduction in
$\Delta_c$ going from $V_{1x}=V_{1y}=3$ to $V_{1x}=1, V_{1y}=5$.
Setting $V_{1x}=0$, however, reduces the gap to $\Delta_c \sim 4$.

In Fig.~\ref{fig6} the signature of phase separation is made more
evident, in the form of a double valued dependence of density on
chemical potential $\mu=dE/d\rho$.  In other words, the chemical
potential $\mu$ is not a monotonically increasing function of density
$\rho$.  The usual Maxwell construction can be used to ascertain the
relative sizes of the superfluid and solid regions in this situation.
The instability of the checkerboard supersolid in the isotropic case
is known\cite{batrouni00}.  Figures~\ref{compare-EanisoV} and
\ref{fig6} establish that this instability persists in the anisotropic
case, except in the extreme case of vanishing interchain interaction,
$V_{1x}=0$.  However, as seen in Fig.~\ref{fig1}, the structure factor
falls off rapidly with doping in this situation, so that long range
density (solid) order is no longer present.  Thus the absence of phase
separation does not imply the existence of a supersolid.

\begin{figure}[htp]
\epsfig{figure=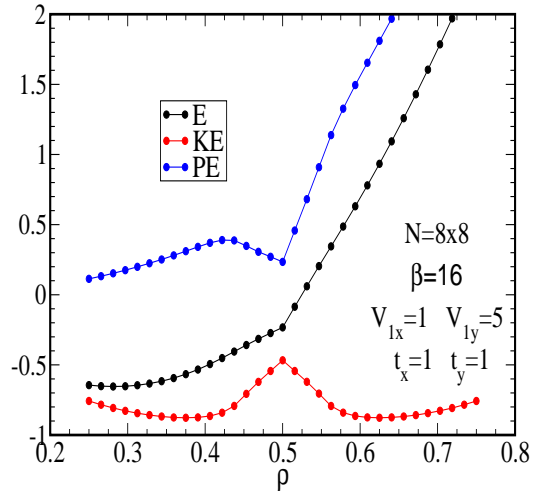,width=8.0cm,height=8.0cm,angle=-90,clip}
\caption{ \label{EKEPEtxty1Vx1Vy5v2} (Color online)
Kinetic, potential, and total
  energies as functions of the filling $\rho$.  The kinetic energy is
  minimized (in absolute value) at half-filling, as expected for an
  insulating solid.  Phase separation appears to be driven primarily
  by the density dependence of the potential energy, whose decrease from
  a maximum at $\rho \approx 0.43$ results in the negative
  curvature in the total energy.  }
\end{figure}

Figure \ref{EKEPEtxty1Vx1Vy5v2} shows the individual components of the
energy.  At $\rho=0.5$ the system is in an insulating checkerboard
solid phase, and has a corresponding minimum in the absolute value of
the kinetic energy.  The potential energy at first increases from the
low density limit, reflecting the increased likelihood of adjacently
occupied sites.  At $\rho \approx 0.43$, however, the potential energy
decreases, despite a rising number of particles.  Presumably this
reflects the onset of density order: a coherent pattern of occupation
in which adjacent sites are rarely occupied.  Evidently, the decrease
of $\langle n_{i} n_{j} \rangle$ for near neighbor sites $
  ij$ associated with solid formation overwhelms the `trivial'
increase $\langle n_{i} n_{j} \rangle \propto \langle n_{
  i} \rangle \, \langle n_{j} \rangle$ expected for noninteracting
bosons.

\subsection{Hopping Anisotropy}

\begin{figure}[htp]
\epsfig{figure=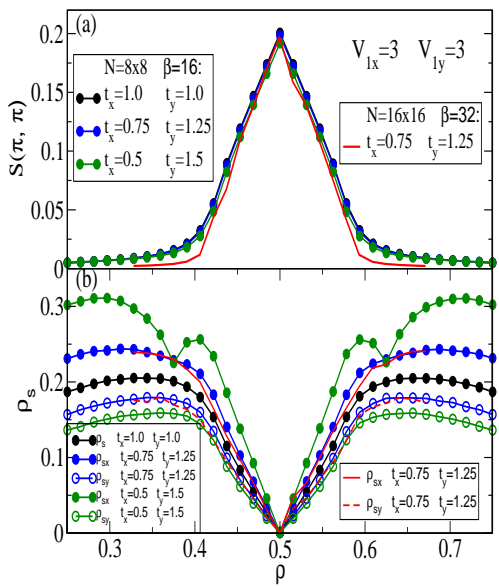,width=8.0cm,height=8.0cm,angle=-90,clip}
\caption{ \label{fig8} (Color online) Checkerboard structure factor
  $S(\pi,\pi)$ (top), along with superfluid densities $\rho_{sx}$ and
  $\rho_{sy}$ in the $\hat x$ and $\hat y$ directions (bottom).
  $S(\pi,\pi)$ is large in the vicinity of half-filling and decreases
  with doping.  It is almost unchanged by hopping anisotropy.  The
  superfluid density vanishes at half-filling, and rises with doping.
  Beginning with $\rho_{sx}=\rho_{sy}$ for $t_x=t_y$, the superfluid
  density in the $\hat x$ direction shifts upward, and that in the
  $\hat y$ direction downward, as $t_x/t_y$ decreases. Results for
  16x16 lattices at $\beta=32$ show that finite spatial lattice and
  temperature effects are minimal. For the largest hopping anisotropy,
  $\rho_{sx}$ has a dip at fillings $\rho=3/8, 5/8$.  This will be
  discussed in a subsequent work.}
\end{figure}

Having explored the possibility of supersolids caused by anisotropy in
the near neighbor interaction we turn to the analogous question when
$t_x \neq t_y$.  Figure~\ref{fig8} shows the checkerboard structure
factor $S(\pi,\pi)$ and superfluid densities $\rho_{sx}$ and
$\rho_{sy}$ in the $\hat x$ and $\hat y$ directions.  The structure factor is
peaked at half-filling but remains relatively large in some interval
surrounding that density.  In general $\rho_{sx} >
\rho_{sy}$, the superflow in the $\hat x$ direction shows a dip at fillings
$\rho=3/8$ and $\rho=5/8$.  This dip becomes more pronounced as the
hopping anisotropy increases.  We will address this issue, which
requires some challenging QMC simulations and which is not directly
related to the questions being addressed here\cite{crepin11}, in a
later paper.

\begin{figure}[htp]
\epsfig{figure=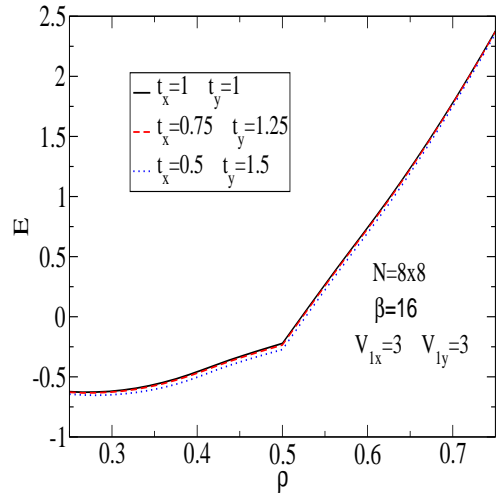,width=8.0cm,height=8.0cm,angle=-90,clip}
\caption{ (Color online)
Energy $E(\rho)$ for three different hopping anisotropies.
  All have the kink which signals the Mott gap, and also negative
  curvature (phase separation) in the vicinity of half-filling.
\label{fig9}
}
\end{figure}

\begin{figure}[htp]
\epsfig{figure=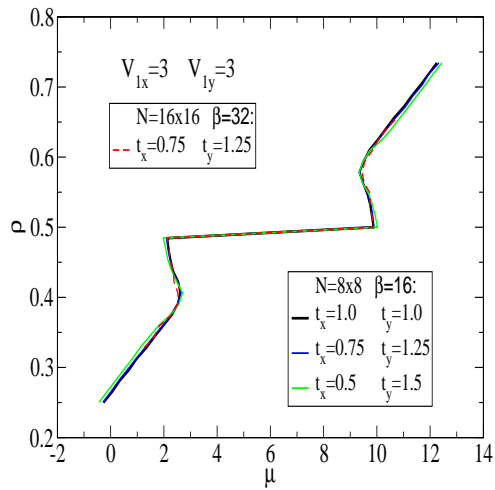,width=8.0cm,height=8.0cm,angle=-90,clip}
\caption{ (Color online) Dependence of filling on chemical potential
  $\rho(\mu)$, extracted from energy data of Fig.~\ref{fig9}.  At
  half-filling, a plateau in $\rho(\mu)$ signals the charge-ordered
  insulator, with adjacent densities showing a thermodynamic
  instability to phase separation. Data for 16x16 lattices at
  $\beta=32$ show that finite spatial lattice and temperature effects
  are minimal both for the size of the Mott gap and the negative
  curvature. The charge gap and negative curvature are largely
  independent of the degree of hopping anisotropy, even though
  $t_{1y}/t_{1x}=3$ is fairly large.  Some of this invariance is due
  to our choice $t_{1x}+t_{1y}=2$, which keeps the noninteracting
  bandwidth fixed.}
\label{fig10}
\end{figure}

The thermodynamics, shown in $E(\rho)$ (Fig.~\ref{fig9}) and
$\rho(\mu)$ (Fig.~\ref{fig10}) show some features similar to those
seen in the anisotropic interaction case. $E(\rho)$ exhibits a kink at
$\rho=0.5$ indicating a jump in the chemical potential and hence a
charge gap, and also negative curvature near $\rho=0.5$ suggesting
phase separation.  $\rho(\mu)$ demonstrates these features even more
directly.  The gap (plateau in $\rho(\mu)$) is unchanged even for a
case when the ratio of hoppings is altered by a factor of three from
$t_x/t_y=1$ to $t_x/t_y=1/3$.  This is in contrast to interaction
anisotropy to which the gap is more sensitive (Figure~\ref{fig6}).

\begin{figure}[htp]
\epsfig{figure=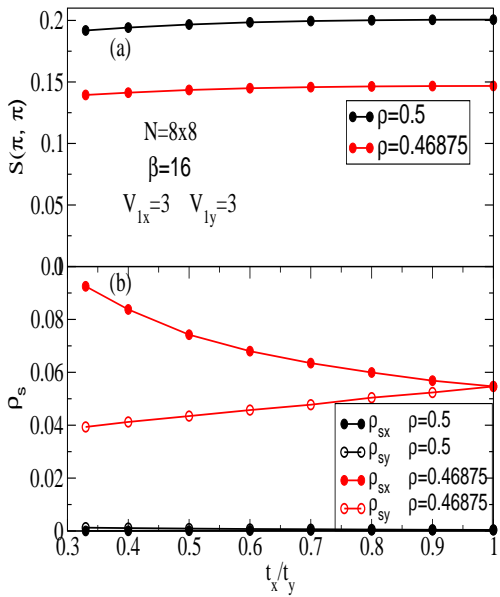,width=8.0cm,height=8.0cm,angle=-90,clip}
\caption{(Color online)
 (a): $S(\pi,\pi)$ vs $t_{x}/t_{y}$ at $\rho=0.5$ and
  $\rho=15/32$. The structure factor is even less sensitive  to
  hopping anisotropy than to interaction anisotropy, as might
  perhaps be expected since charge order is directly driven
  by $V$ as opposed to $t$. (b): $\rho_{sx}$ and $\rho_{sy}$ vs
  $t_{x}/t_{y}$ at $\rho=0.5$ and $\rho=15/32$.
The sum $t_x+t_y=2$ for all data here.}
\label{fig11}
\end{figure}

Figure \ref{fig11} provides quantitative detail on the evolution of
the CDW structure factor and superfluid densities with hopping
anisotropy.  For both densities, the data for $S(\pi,\pi)$ are
quite flat over the range shown, $1/3 < t_x/t_y < 1$.  The superfluid
densities, especially in the $\hat x$ direction show significant
quantitative evolution\cite{foot1}, but qualitatively they are unchanged,
indicating superfluid behavior for density $\rho=15/32$ and insulating
behavior for half-filling $\rho=1/2$.

\section{Conclusions}

The simultaneous presence of diagonal and off-diagonal long range
order, as opposed to a transition between distinct ordered phases, is
by now well-established in many bosonic models.  In this paper we have
shown that this coexistence is thermodynamically unstable in a broad
range of square lattice hard-core Hamiltonians with near neighbor
interactions and hopping parameters which break $x$-$y$ symmetry.
Although our results are for choices in which only hopping or
interactions are anisotropic, we have also studied a few cases where
x/y symmetry is broken in both terms \cite{foot2}.  These more complex
situations also exhibited phase separation as opposed to
supersolidity.  Of course, we cannot exclude the possibility that a
supersolid is stable in some other choice of parameter values.

The motivation for studying such situations is to gain a more complete
understanding of the interplay between real space particle
arrangements and superflow in quantum systems.  For strongly
correlated fermions, this is a central puzzle of unconventional
superconductivity: Do antiferromagnetic (AF) and charge order
(stripes) compete or coexist with pairing?  These issues arise not
only in cuprate materials, which have AF ordering at $q =
(\pi,\pi)$ but also in the more recently discovered iron-pnictides
whose magnetic ordering is at $q = (0,\pi)$: AF in one direction
and ferromagnetic in the other.

Since bosonic systems are more simple to study computationally than
fermionic ones, it would be interesting to extend the present work to
`spin-1/2' bosons\cite{deparny10} where one might have the analogs of
both spin and charge order along with the possibility of
superfluidity.


This work was supported by:
the National Key Basic Research Program of
China Grant No.~2013CB328702;
a CNRS-UC Davis EPOCAL LIA joint research grant;
by NSF-PIF-1005503; DOE SSAAP
DE-NA0001842-0; DOE SciDAC Grant No. DE-FC02-06ER25792
and NSF OISE-0952300.
We thank B. Brummels for useful input.



\begin{thebibliography}{100}
\bibitem{explainSGF} Although this capability is not needed
for the situation studied here, the SGF method is especially
suited to complex models containing general four and six operator
terms in the Hamiltonian.
\bibitem{foot1}  In principle, one might ask why the anisotropy
breaks the symmetry between $\rho_{sx}$ and $\rho_{sy}$ in the
way it does, i.e.~why $\rho_{sx} > \rho_{sy}$ when $t_x < t_y$.
However, it is unclear whether there is a meaningful answer to
this question in this case where the system phase separates.
\end{thebibliography}
\end{document}